\documentclass[twocolumn,floatfix,showkeys]{emulateapj}
\usepackage{latexsym}
\usepackage{epsfig}
\usepackage{amssymb}
\usepackage[latin1]{inputenc}
\usepackage[T1]{fontenc}
\usepackage{amsmath}
\usepackage{pstricks}
\usepackage{graphicx}

\begin{document}
\title{Cosmology of Chameleons with Power-Law Couplings}
\author{David F. Mota \& Hans A. Winther}
\affil{Institute of Theoretical Astrophysics, University of Oslo, Box 1029, 
0315 Oslo, Norway}

\begin{abstract}
In chameleon field theories a scalar field can couple to matter with gravitational strength and still evade local gravity constraints due to a combination of self-interactions and the couplings to matter. Originally, these theories were proposed with a constant coupling to matter, however, the chameleon mechanism also extends to the case where the coupling becomes field-dependent. We study the cosmology of chameleon models with power-law couplings and power-law potentials. It is found that these generalized chameleons, when viable, have a background expansion very close to $\Lambda \rm CDM$, but can in some special cases enhance the growth of the linear perturbations at low redshifts. For the models we consider it is found that this region of the parameter space is ruled out by local gravity constraints. Imposing a coupling to dark matter only, the local constraints are avoided, and it is possible to have observable signatures on the linear matter perturbations.
\end{abstract}
\maketitle
%


\section{Introduction}
The origin of dark energy (DE) responsible for the cosmic acceleration remains a mystery. A host of independent observations have supported the existence of dark energy over the past decade, however, no strong evidence was found yet implying that dynamical DE models are better than a cosmological constant (see e.g. \cite{wood} for results on observation of dark energy and \citep{copeland,durrer} for theoretical overviews). The first step towards understanding the origin of DE would be to detect some clear deviation from the $\Lambda\rm CDM$ model observationally and experimentally.

Scalar fields (quintessence) are natural DE candidates, with an equation of state different from $-1$. In quintessence models \citep{wang1,zlatev} the scalar field is slowly rolling down its potential, its energy density is dominated by the potential energy and almost remaining constant provided that the potential is flat enough. However, this means that the mass of the scalar field is in general very light and as a result the scalar field almost does not cluster so that its effects in cosmology are mainly on the (modified) background expansion rate.

However, a certain class of theories have been proposed, in which the scalar field(s) properties depend on the environment: these are the class of chameleon field theories, proposed in \cite{Khoury:2003rn}, that employed a combination of self-interactions of the scalar-field and couplings to matter to avoid the most restrictive of the current bounds. In the models that they proposed, a scalar field couples to matter with gravitational strength, in harmony with general expectations from string theory, whilst, at the same time, remaining relatively light on cosmological scales. Originally, the chameleon was proposed with a constant coupling to matter, however, it was shown in \cite{Brax:2010kv} that the chameleon mechanism extends to the case where the coupling becomes field-dependent. The chameleon mechanism can also be present in other types of models like shown in \cite{kchameleon} for the case of k-essence. Because chameleons are allowed to couple strongly to matter and at the same time exhibit a long ranged fifth-force in space, they could leave a strong signature on the growth of the matter perturbations. Chameleon models also have signatures in local gravity experiments which can be within reach of near-future experiments \citep{Khoury:2003rn,Brax:2010xx,local2}.

The modified evolution of the matter density perturbations can provide an important tool to distinguish generally modified gravity DE models, from DE models inside GR like the $\Lambda\rm CDM$ model \citep{de1,de2,de3,de4,de5,de6,de7,barrow,de8,de9,de10,de11,de12,de13}. In fact the effective gravitational constant $G_{\rm eff}$ which appears in the source term driving the evolution of matter perturbations can change significantly relative to the gravitational constant $G$. A useful way to describe the perturbations is to write the growth function $f = \frac{d\log\delta_m}{\log a}$ as $f = \Omega_m(z)^{\gamma}$ where $\Omega_m$ is the density parameter of non-relativistic matter (baryonic and dark matter) \citep{a1,a2}. One has $\gamma\approx 0.55$ at low redshifts in the $\Lambda\rm CDM$-model \citep{wang,linder1,linder2}. While $\gamma$ is quasi-constant in standard (non-interacting) DE models inside GR, this needs not be the case in modified gravity models. An additional important point is whether $\gamma$ can exhibit scale dependence (dispersion). When this happens the resulting matter power spectrum acquires an additional scale dependence which is not found in $\Lambda\rm CDM$.

In \cite{2010arXiv1010.3769G} a chameleon model, with a constant coupling, was constructed where the present value of the growth index $\gamma$ can be as small as $\gamma = 0.2$ together with a significant redshift dependence. This allows to clearly discriminate this model from $\Lambda\rm CDM$.

In this paper, we investigate the cosmological properties of chameleon models where the matter-coupling is a power-law function of the scalar-field. This article is divided in 5 main sections. In section 2 we give a short review of chameleon models and define the models we will be looking more closely at. In section 3 we discuss the background evolution. In section 4 we discuss local and cosmological constraints on our models, and in section 5 we consider the evolution of the matter perturbations in our models.


\section{Chameleon Theories}
In this section we review the basic properties of chameleon models. We start by considering the scalar-tensor theory described by the action
\begin{align}\label{cham_action_1}
S = \int dx^4 \sqrt{-g}\left[\frac{RM_{\rm pl}^2}{2} - \frac{1}{2}(\partial\phi)^2 - V(\phi)\right] + S_{\text{m}}(\tilde{g}_{\mu\nu}^{(i)},\psi_i)
\end{align}
where $M_{\rm pl} = \frac{1}{\sqrt{8\pi G}}$ is the reduced Planck scale, $g$ is the determinant of the metric $g_{\mu\nu}$, $R$ is the Ricci-scalar with respect to $g_{\mu\nu}$ and $\psi_i$ are the different matter-fields (including radiation). The matter field $i$ couples to the Jordan-frame metric $\tilde{g}_{\mu\nu}^{(i)}$ which is related to the Einstein-frame metric $g_{\mu\nu}$ via a conformal rescaling on the form
\begin{equation}\label{conformal_coupling}
\tilde{g}_{\mu\nu}^{(i)} = e^{2\beta_i(\phi)}g_{\mu\nu}
\end{equation}
In the following we focus, for simplicity, on the case where all the matter-fields couple to $\phi$ with the same $\beta_i(\phi)\equiv \beta(\phi)$. Thus, a general model of this kind is then uniquely specified by stating the two functions $V(\phi)$ and $\beta(\phi)$. 

A variation of the action Eq.~(\ref{cham_action_1}) with respect to $\phi$ yields the field-equation
\begin{equation}
\square\phi = V_{,\phi} - \beta,_{\phi}(\phi) e^{4\beta(\phi)}\tilde{g}_{\mu\nu}^{(i)}\tilde{T}^{\mu\nu}_{(i)}
\end{equation}
where $\tilde{T}^{\mu\nu} = \frac{2}{\sqrt{-g}}\frac{\partial \mathcal{L}_m}{\partial g_{\mu\nu}}$ is the energy-momentum tensor of the matter fields. The energy-momentum tensor for radiation vanishes meaning that chameleons do not couple directly to photons. However, a coupling to photons, which have some interesting observable signatures  \citep{cp1,cp2,cp3,solarcham},  can be introduced by modifying the electromagnetic field strength $F_{\mu\nu}^2\to e^{\beta_{\gamma}(\phi)}F_{\mu\nu}^2$.

In the perfect fluid approximation, we have $\tilde{g}_{\mu\nu}\tilde{T}^{\mu\nu} = -\tilde{\rho}_m$ where $\tilde{\rho}_m$ is the Jordan-frame energy density of non-relativistic matter. The density $\tilde{\rho}_m$ is conserved with respect to the Jordan-frame metric $\tilde{g}_{\mu\nu}$. In the Einstein-frame the density $\rho_m \equiv \tilde{\rho}_m e^{3\beta(\phi)}$ is both conserved and $\phi$-independent. The field equation, in the Einstein-frame, can be written
\begin{align}
\begin{array}{rcl}
\square\phi &=& V_{\rm eff,\phi}\\
V_{\rm eff}(\phi) &=& V(\phi) + \rho_m e^{\beta(\phi)}
\end{array}
\end{align}
where $V_{\rm eff}$ is the effective potential. In the following, the quantity $\beta,_{\phi}$ will be referred to as the coupling and $\beta(\phi)$ as the coupling-function. We will also use the notation $A,_{\phi} \equiv \frac{dA}{d\phi}$ and $A,_{\phi_a} \equiv \left.\frac{dA}{d\phi}\right|_{\phi=\phi_a}$.


\subsection{Chameleon thin-shell mechanism}
We will in this section give a short review of the chameleon thin-shell mechanism in which these theories can have a strong matter coupling and still be viable.

The coupling Eq.~(\ref{conformal_coupling}) of $\phi$ to matter leads to the introduction of a fifth-force in nature. In linear theories of massive scalar fields the superposition principle holds, meaning that the larger a massive body is the stronger the fifth-force becomes. Chameleon theories behave quite opposite: In situations where massive bodies are involved, the chameleon field is trapped inside the bodies and its influence on other bodies is only due to a thin shell close to the surface of the bodies. This leads to a shielded fifth-force which becomes hard to detect.

In order to have a chameleon mechanism in a theory the following conditions must be satisfied
\begin{itemize}
	\item For a given density $\rho_0$ the effective potential $V_{\rm eff} = V(\phi) + \rho_0 e^{\beta(\phi)}$ must exhibit a minimum $\phi_{\rm min}$.
	\item The mass, $m_{\rm min} \equiv \sqrt{V_{\rm eff,\phi\phi}(\phi_{\rm min})}$, of small oscillations about this minimum must be a real increasing function of $\rho_0$.
\end{itemize}
and puts the following constraints on the potential $V(\phi)$ and coupling-function $\beta(\phi)$:
\begin{align}
	{\rm i)}~V,_{\phi}\beta,_{\phi} < 0,~~~ {\rm ii)}~V,_{\phi\phi\phi}V,_{\phi}>0,~~~ {\rm iii)}~\beta,_{\phi\phi\phi}\beta,_{\phi}>0
\end{align}
Note that the second condition above is required if we want the chameleon to exhibit a long-ranged force in the vacuum of space. The final requirement, which is the most important,  is that there exists a chameleon thin-shell mechanism in which the fifth-force, for sufficiently large bodies, becomes suppressed relative to the gravitational force.

To see how this mechanism works in practice, we consider a spherical object with constant density $\rho_c$, radii $R_c$ and mass $M_c$ embedded in a background of homogeneous density $\rho_b\ll \rho_c$. In such a static spherical symmetric space time, the field equation reads
\begin{align}
	\frac{d^2\phi}{dr^2}+\frac{2}{r}\frac{d\phi}{dr} = V_{\rm eff,\phi} = \left\{\begin{array}{l}V,_{\phi}+\rho_c\beta,_{\phi}e^{\beta(\phi)}~~r<R_c\\V,_{\phi}+\rho_b\beta,_{\phi}e^{\beta(\phi)}~~r>R_c\end{array}\right.
\end{align}
The effective potential have two minimums at $\phi=\phi_c$ and $\phi=\phi_b$ satisfying
\begin{align}
	V,_{\phi}(\phi_c)+\beta,_{\phi}(\phi_c)\rho_ce^{\beta(\phi_c)}=0\\
	V,_{\phi}(\phi_b)+\beta,_{\phi}(\phi_b)\rho_be^{\beta(\phi_b)}=0
\end{align}
The $\phi$-mediated fifth-force per unit mass is given by
\begin{align}\label{force_law}
	\vec{F}_{\phi} = - \beta,_{\phi}\vec{\nabla}\phi~~~\text{for}~~~r>R_c
\end{align}
In the following we will consider the simplest case $\beta(\phi) = \frac{Q\phi}{M_{\rm pl}}$, in which the coupling $\beta,_{\phi} = Q/M_{\rm pl}$ is constant, together with an arbitrary runaway potential like e.g. $V(\phi) = \frac{M^{n+4}}{\phi^n}$.

For this case, the field profile outside the body have been derived in \cite{Khoury:2003rn} and reads
\begin{align}
\phi = \phi_b - \frac{Q_{\rm eff}}{4\pi M_{\rm pl}}\frac{M_c}{r}e^{-m_b r}	
\end{align}
where $m_b = \sqrt{V_{\rm eff,\phi\phi}(\phi_b)}$ is the mass of the field in the background and $Q_{\rm eff}$ is the effective coupling. The value for $Q_{\rm eff}$ depends on the value of the thin-shell factor
\begin{align}
	\epsilon_{\rm th} \equiv \frac{\phi_b-\phi_c}{6Q M_{\rm pl}\Phi_c}
\end{align}
where $\Phi_c$ is the Newtonian gravitational potential. When the body is small in the sense that $\epsilon_{\rm th}>1$ the field acts like a linear scalar field and $Q_{\rm eff} = Q$. 

For large bodies, $\epsilon_{\rm th}\ll 1$, the field will be stuck at the minimum $\phi_c$ inside the body, and the only changes in $\phi$ takes place in a thin-shell close to the surface. We say that the body has developed a thin-shell.

In this case we find $Q_{\rm eff} = 3 Q \epsilon_{\rm th}$. Thus, the coupling strength is suppressed relative to $Q$. More accurate formulas for $Q_{\rm eff}$ when $\epsilon_{\rm th}\lesssim 1$ can be found in \cite{Tamaki:2008mf}.

The amplitude of the fifth-force on a test-particle of unit mass outside the body is
\begin{align}
	F_{\phi} = 2 Q Q_{\rm eff} \frac{GM_c}{r^2}~~~\text{for}~~~ r<m_b^{-1}
\end{align}
For a sufficiently large body we have $2 Q Q_{\rm eff}\ll 1$, and the resulting fifth-force is suppressed relative to the gravitational force. When test-masses used in local gravity experiments have thin-shells, the experimental bounds are easily evaded.

For more general coupling-functions $\beta(\phi)$ the analysis above can be much more complicated due to the coupling $\beta,_{\phi}$ in Eq.~(\ref{force_law}) being field-dependent, see \cite{Brax:2010kv}). We present a derivation of the thin-shell solution for the power-law coupling $\beta(\phi) = \left(\frac{\lambda\phi}{M_{\rm pl}}\right)^m$ with $m>1$ in section 4.


\subsection{Our models}
We will look more closely on the following models, in which the cosmological constant is part of the potential:
\begin{align}\label{modela}
	{\rm Model~A:} \left\{\begin{array}{l}V(\phi) = M^4\left[1+ \left(\frac{M}{\phi}\right)^n\right]\\ \beta(\phi) = \left(\frac{\lambda\phi}{M_{\rm pl}}\right)^m\end{array}\right.
\end{align}
with $n>0$ and $m\geq 1$. This model is a generalization of the original chameleon model ($m=1$) introduced in \cite{Khoury:2003rn}.
\begin{align}\label{modelb}
	{\rm Model~B:} \left\{\begin{array}{l}V(\phi) = M^4\left[1+\left(\frac{\phi}{M}\right)^n\right]\\ \beta(\phi) = \left(\frac{\lambda H_0}{\phi}\right)^m\end{array}\right.
\end{align}
with $n>2$, $m>0$ and $H_0$ being the current Hubble parameter. This model was introduced, and a wide range of local gravity constraints was calculated, in \cite{Brax:2010kv}.

In order for the chameleon to act as dark energy we need to choose $M^4\sim \Lambda$ ($M\sim 10^{-3}$eV). Note that this value is also required in order to satisfy local gravity constraints. Thus, the fine-tuning problem of the cosmological constant is also present in these models.


\section{Background Evolution}
In this section we discuss the background cosmological evolution of chameleon models.

We consider a flat Friedmann-Lemaitare-Robertson-Walker (FLRW) background metric
\begin{align}\label{flrw}
	ds^2 = -dt^2 + a(t)^2(dx^2+dy^2+dz^2)
\end{align}
The corresponding background equations are given by
\begin{align}
&3H^2M_{\rm pl}^2 = \rho_m e^{\beta(\phi)} + \rho_r + \frac{1}{2}\dot{\phi}^2 + V(\phi)\\
&\dot{\rho}_m+3H\rho_m = 0\\
&\dot{\rho}_r+4H\rho_r = 0
\end{align}
The field equation for $\phi$ becomes
\begin{align}\label{cosm_phi_eq}
\begin{array}{l}
\ddot{\phi} + 3H\dot{\phi} + V_{\text{eff},\phi} = 0\\
V_{\text{eff}} = V(\phi) + \rho_m e^{\beta(\phi)}
\end{array}
\end{align}
We also introduce the density parameters
\begin{align}
\begin{array}{l}
\Omega_r = \frac{\rho_r}{3H^2M_{\rm pl}^2}\\
\Omega_m = \frac{\rho_m e^{\beta(\phi}}{3H^2M_{\rm pl}^2}\\
\Omega_{\phi} = \frac{V + \frac{1}{2}\dot{\phi}^2}{3H^2M_{\rm pl}^2}
\end{array}
\end{align}
The background evolution of the original chameleon model introduced in \cite{Khoury:2003rn}, which corresponds to Model A Eq.~(\ref{modela}) with $m=1$, was thoroughly discussed in \cite{Brax:2004qh}. We will in the next section show that the chameleon behaves very similar in the general setting. There are however some important differences.


\subsection{Attractor solution}
We show the existence of an attractor solution where the chameleon follow the minimum of its effective potential $\phi = \phi_{\rm min}(t)$ as long as the condition $m_{\phi} \gg H$ is satisfied.

Suppose the field is at the minimum at some time $t_i$. Then a time later due to the red shifting of the matter density the minimum $\phi_{\rm min}$ has moved to a slightly larger value (or smaller, depending on the form of the coupling). The characteristic timescale for this evolution is the Hubble time $1/H$. Meanwhile the characteristic timescale of the evolution of $\phi$ is given by $1/m_{\phi}$. When $m_{\phi}\ll H$ the response-time of the chameleon is much larger than $1/H$, the chameleon cannot follow the minimum and starts to lag behind. But if $m_{\phi}\gg H$ then the response-time of the chameleon is much smaller than $1/H$, the chameleon will adjust itself and adiabatically start to oscillate about the minimum. This can also be seen from the analogy of Eq.~(\ref{cosm_phi_eq}) with a driven harmonic oscillator
\begin{equation}
\ddot{x}+2\zeta\omega\dot{x} + \omega^2x =0
\end{equation}
This equation will have a solution which oscillates with a decreasing amplitude as long as $\zeta < 1$. When the field is close to the minimum we can approximate
\begin{equation}
\ddot{\phi}+3H\dot{\phi} + m_{\phi}^2(\phi-\phi_{\rm min}) =0
\end{equation}
The condition $\zeta<1$ reduces to $\frac{2m_{\phi}}{3H} > 1$, and we will require that $m_{\phi}^2 \gg H^2$ is satisfied from the early universe and until the present era.


\subsection{Reaching the attractor}
We discuss how the chameleon acts in the early universe, and how the field converges to the attractor. For simplicity we will only focus on the case where $\dot{\phi}_{\rm min} < 0$ (Model B Eq.~(\ref{modelb})), the case $\dot{\phi}_{\rm min}>0$ (Model A Eq.~\ref{modela})) is analogous. 

If we release the field at some time $t_i$, at some initial value $\phi_i$, we can have two cases:


\subsubsection{Undershooting: $\phi_i \gg \phi_{\rm min}(t_i)$}
In this case the field equation can be approximated by
\begin{equation}\label{cosm_undershoot_app}
\ddot{\phi}+3H\dot{\phi} \approx -V,_{\phi}
\end{equation}
which is the same equation as in quintessence. The driving term will dominate over the friction term when $V_{,\phi\phi}\gg H^2$ and will start driving the field down towards the minimum. If this condition is not satisfied, the field will be fixed at $\phi_i$ until the Hubble factor has had time to be sufficiently redshifted allowing the field to start rolling down the potential.

When the field starts to roll it drops to, and go past, $\phi=\phi_{\rm min}$. Here the approximation Eq.~(\ref{cosm_undershoot_app}) cannot be used anymore, but the field will usually have to much kinetic energy to settle at the minimum and will be driven past it. The further below the minimum the field is driven, the larger the factor $\beta,_{\phi}\rho_m$ becomes. Even though $\rho_m$ is very small in the radiation era it will eventually kick in and drive the field up again. More importantly, we will also have a contribution from the decoupling of relativistic matter which will be discussed in the next section. This will make the field oscillate around the minimum, and as long as $m_{\phi}^2/H^2\gg 1$ the amplitude of the oscillations will be damped, making sure that the field quickly converges to the attractor. This can be showed explicitly as done in \cite{Brax:2004qh}, the derivation found there is general and applies to our case as well.


\subsubsection{Overshooting: $\phi_i \ll \phi_{\rm min}(t_i)$}
In this case the potential term $V,_{\phi}$ can be ignored and the $\phi$-equation becomes
\begin{equation}
\ddot{\phi}+3H\dot{\phi} \approx \beta,_{\phi}T^{\mu}_{\mu}
\end{equation}
where we have restored the trace of the energy-momentum (EM) tensor. In the radiation-dominated era this trace is very small since radiation does not contribute to the trace, and the field will be frozen at its initial value. As the universe expands and cools the different matter-species decouple from the radiation heat bath when the temperature is of the same order as the mass of the matter-particles. This gives rise to a trace-anomaly where the trace of the EM-tensor gets non-zero for about one e-fold of expansion leading to a 'kick' in the chameleon pushing it to larger field-values. This trace can be written for a single matter-species, see \cite{Brax:2004qh}, as
\begin{equation}
T^{\mu (i)}_{\mu} = -\frac{45}{\pi^4}H^2M_{\rm pl}^2\frac{g_i}{g*(T)}\tau(m_i/T)
\end{equation}
where $g_*$ is the effective number of relativistic degrees of freedom, $g_i$, $T_i$ and $m_i$ is the degrees of freedom the temperature and the mass of species $i$ respectively. The $\tau$-function is given by
\begin{equation}
\tau(x) = x^2\int_x^{\infty}\frac{\sqrt{y^2-x^2}dy}{e^y\pm 1}
\end{equation}
and $\pm$ refers to bosons and fermions respectively.

\begin{figure}[htbp]
	\centering
		\includegraphics[width=\columnwidth]{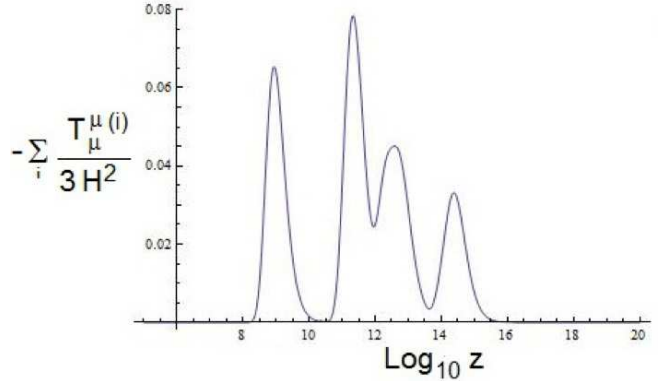}
	\caption{The trace of the EM-tensor, $\Omega_{\text{m eff}} = -T^{\mu}_{\mu}/(3H^2M_{\rm pl}^2)$, in the radiation dominated era for all the different matter species decoupling from the radiation heat bath.}
	\label{fig:taufunction}
\end{figure}

See Fig.~(\ref{fig:taufunction}) for a plot of $T^{\mu}_{\mu}/(3H^2M_{\rm pl}^2)$ in the radiation era.
The plot shows that each kick contributes to the field equation approximately as an effective matter-density $\Omega_{\text{m eff}}\sim \mathcal{O}(0.01)$. By using a delta-function source as the kick we can show that the result is to push the field up a distance
\begin{equation}
|\Delta\phi| = \mathcal{O}\left(|\beta,_{\phi_i}|M_{\rm pl}^2\frac{g_i}{g*(m_i)}\right)
\end{equation}
where $\phi_i$ is the field-value before the kick sets in. When $\beta,_{\phi} = {\rm constant}$, the field will be kicked almost the same amount each time a new species freezes out. But when $\beta,_{\phi} \not= {\rm constant}$ the smaller the initial value $\phi_i$ the more effective the kick is in bringing it back up. When above the minimum the resulting kicks will be balanced by the term $V,_{\phi}$ which drives the field down again making the field oscillate above the minimum before eventually settling down.

\begin{figure}[htbp]
	\centering
		\includegraphics[width=\columnwidth]{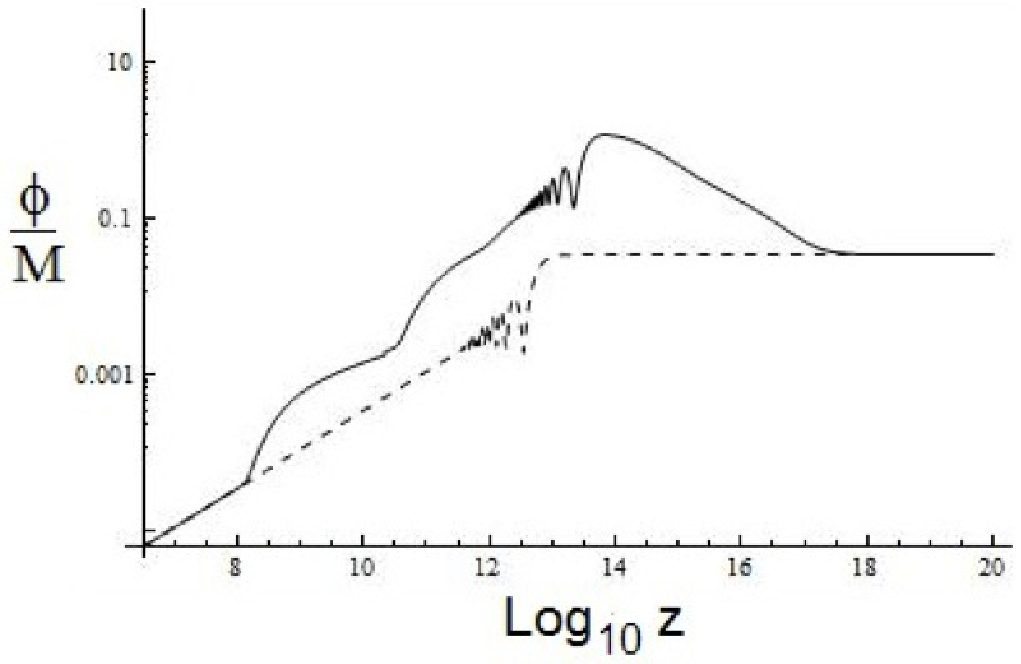}\\
		\includegraphics[width=\columnwidth]{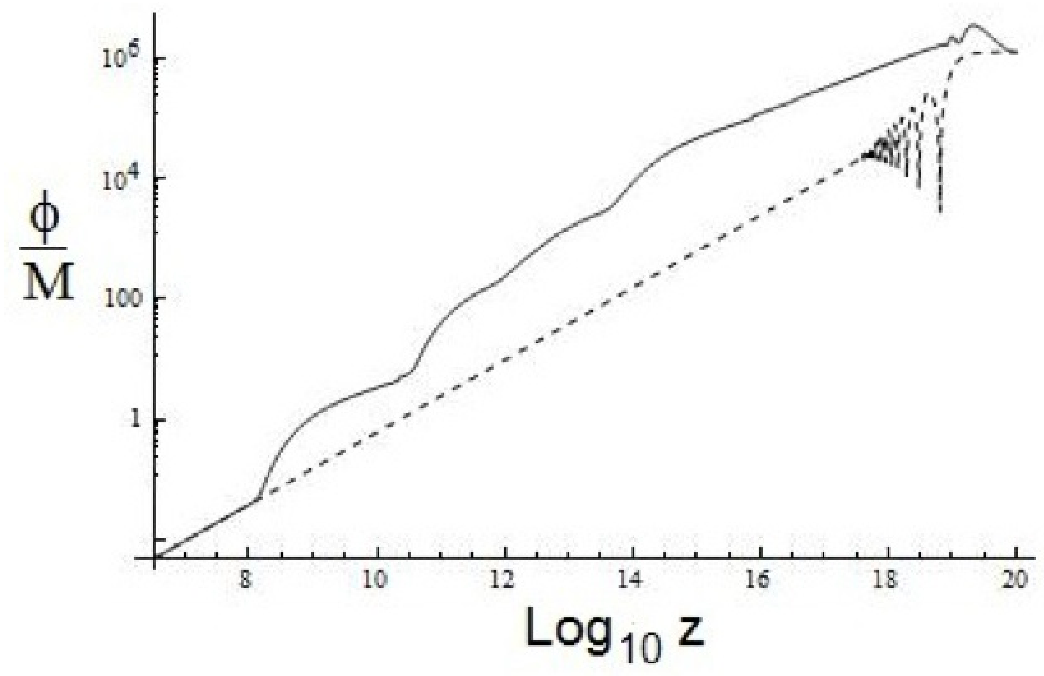}
	\caption{The evolution of $\phi$ in the early universe as a function of the redshift $z$ for two types of initial conditions: $\phi_i \gg \phi_{\rm min}$ (above) and $\phi_i \ll \phi_{\rm min}$ (below). In this example $V(\phi) = M^4+\phi^4,~~\beta(\phi)=\frac{H_0}{\phi}$. The dashed (full) line corresponds to the solution where we neglect (include) the kicks. The kicks-solution does not reach the minimum until $z\approx 10^8$, but due to the large mass of the field it starts to follow it right away.}
	\label{fig:cosmology_n4_M30_L1}
\end{figure}

Due to this mechanism, the chameleon can have initial values far below the minimum and still be able to get close to the attractor relative quickly. The initial value will of course depend on the how the chameleon behaves under inflation. If the chameleon couples to the inflaton and sits at the minimum at the onset of inflation, then after the inflaton decays to reheat the universe the density of matter-species coupled to the chameleon will decrease rapidly since most of the energy will go to radiation. This will lead to a release of the field at a value well above the minimum, where the undershoot solution applies. As long as $m_{\phi}\gg H$, the field will typically settle at the minimum before the onset of the Big Bang Nucleosynthesis (BBN). See Fig.~(\ref{fig:cosmology_n4_M30_L1}) for a typical evolution of $\phi$ in the early universe with and without the inclusion of the kicks.

\subsection{Dynamics of $\phi$ along the attractor}
When the field follows the attractor $\phi \approx \phi_{\rm min}$, we have $V_{\text{eff},\phi} \approx 0$. Taking the time-derivative yields
\begin{equation}\label{phi_min_dot}
\dot{\phi} \approx -3H\frac{V,_{\phi}}{m_{\phi}^2}
\end{equation}
for $\beta(\phi)\ll 1$. We further find
\begin{equation}
\frac{\dot{\phi}^2}{2V}=\frac{9H^2}{2m_{\phi}^2}\frac{1}{\Gamma}
\end{equation}
where
\begin{equation}
\Gamma = \frac{Vm_{\phi}^2}{V_{,\phi}^2}
\end{equation}
As long as $\Gamma > 1$, the field will be slow-rolling whenever the condition $m_{\phi}^2\gg H^2$ is satisfied.

\begin{figure}[htbp]
	\centering
		\includegraphics[width=\columnwidth]{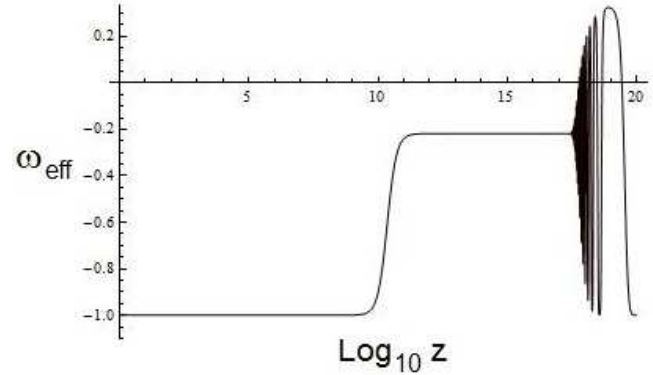}
	\caption{The effective equation of state for the chameleon when $V(\phi) = M^4+\phi^4,~~\beta(\phi)=\frac{H_0}{\phi}$. We have $\omega = -\frac{m+n}{n} = -0.2$ during the period before BBN. After the transition to $\phi < M$, the chameleon behaves like a cosmological constant with $\omega=-1$. The oscillations in $\omega_{\rm eff}$ comes from the field oscillating around the minimum before eventually settling down.}
	\label{fig:omegaeff}
\end{figure}

The equation of state for a minimal coupled scalar field (quintessence) is given by $\omega_{\phi} = \frac{\dot{\phi}^2-2V}{\dot{\phi}^2+2V}$. Since the chameleon is not minimal coupled, the time evolution of $\rho_{\phi}$ must be computed directly from $\dot{\rho_{\phi}}/\rho_{\phi} = -3H(1+\omega_{\text{eff}})$. Using Eq.~(\ref{phi_min_dot}), we find
\begin{equation}
\omega_{\text{eff}} = -1 + \frac{1}{\Gamma}
\end{equation}
From this equation we see that the chameleon acts as a dark energy fluid as long as $\frac{3}{2} < \Gamma$, and its only for models where $\Gamma\sim \mathcal{O}(1)$ where we can have a significant deviation from $\omega=-1$. See Fig.~(\ref{fig:omegaeff}) for a typical evolution of $\omega_{\rm eff}$.

Model A Eq.~(\ref{modela}) yields
\begin{align}
	\Gamma = 1+\frac{m}{n} + \frac{(n+m)}{n}\left(\frac{\phi_{\rm min}}{M}\right)^n
\end{align}
The transition from $\phi_{\rm min} < M$ to $\phi_{\rm min} > M$ takes place for a redshift
\begin{align}
	(1+z) = \left[\frac{n\Omega_{\phi 0}}{m\Omega_{m0}}\left(\frac{M_{\rm pl}}{M\lambda}\right)^{m}\right]^{1/3} \approx 10^{10m}\lambda^{-m/3}
\end{align}
which for typical values of $(m,\lambda)$ is before the time of BBN ($z\approx 10^9)$. This means that in the background today we have $\phi \gg M$ and therefore $V(\phi) \approx M^4$ and $\Gamma \gg 1$.

For Model B Eq.~(\ref{modelb}) we find
\begin{align}
	\Gamma = 1 + \frac{m}{n} + \frac{(n+m)}{n}\left(\frac{M}{\phi_{\rm min}}\right)^n
\end{align}
The transition from $\phi_{\rm min} > M$ to $\phi_{\rm min} < M$ takes place for a redshift
\begin{align}
	(1+z) = \left[\frac{m\Omega_{\phi 0}}{n\Omega_{m0}}\left(\frac{M}{H_0\lambda}\right)\right]^{\frac{m}{3}} \approx 10^{10m}\lambda^{-\frac{m}{3}}
\end{align}
where we have used $\frac{M}{H_0}\sim \frac{M_{\rm pl}}{M} \sim 10^{30}$. For typical values of $(m,\lambda)$ we find a redshift which also is before the time of BBN. Today this translates into $\phi \ll M$ giving $V(\phi) \approx M^4$ and $\Gamma \gg 1$. 

Thus the models we consider here, will have a background evolution very close to that of $\Lambda \rm CDM$.


\subsection{Statefinder parameters}
The statefinder diagnostics, introduced in \cite{Sahni:2002fz}, can be a useful tool for distinguishing different DE models. The statefinder parameters are defined by
\begin{align}
	r = \frac{\ddot{a}}{aH^3},~~~ s = \frac{r-1}{3(q-1/2)}
\end{align}
where $q = -\frac{\ddot{a}}{aH^2}$ is the deceleration parameter. Upon defining $h \equiv H^2$, it follows
\begin{align}
	q = -1 + \frac{h'}{2h},~~~r = 1+\frac{h''}{2h}-\frac{3h'}{2h}
\end{align}
where a prime denoted a derivative relative to $x=-\log(a)$. $\Lambda\rm CDM$, neglecting the contribution from radiation, corresponds to the fixed point $(r,s)=(1,0)$. For a general chameleon (in units of $M_{\rm pl}\equiv 1$) we find
\begin{align}
	&r =  1 + \frac{3}{2}(\phi')^2 -  \frac{3}{2}\Omega_m\beta,_{\phi}\phi' +  \frac{V,_{\phi}}{h}\phi'\\
	&s = \left(\frac{2}{3}\frac{V,_{\phi}}{V}\phi' + \frac{1}{3}\frac{\rho_m\beta,_{\phi}}{V}-\frac{(\phi')^2h}{V}\right)\frac{1}{1- \frac{(\phi')^2h}{2V}}
\end{align}
When the chameleon is slow rolling along the minimum of the effective potential, that is $\phi ' \approx \frac{3V,_{\phi}}{m_{\phi}^2}$ and $m_{\phi}^2 \gg H^2$, we can simplify the above equations to
\begin{align}
	r - 1 &\simeq  \frac{27}{2}\frac{1}{\Gamma}\Omega_{\phi} \\
	s &\simeq -\frac{3}{\Gamma}
\end{align}
When $\Gamma \gg 1$ we find $r\approx 1$ and $s\approx 0$. It is only for $\Gamma\sim \mathcal{O}(1)$ that we can have an (observable) deviation from $\Lambda \rm CDM$. For the models considered here, $\Gamma \gg 1$ and the statefinders will be the same as in $\Lambda \rm CDM$. See \cite{2010arXiv1010.3769G} for a viable chameleon model where the statefinder parameters deviates significantly from $\Lambda \rm CDM$.


\section{Constraints}

\subsection{Local gravity constraints}
Experimental tests of general relativity in the solar system, see e.g. \cite{Will:93}, and searches for a fifth-force in nature \citep{eotwash,eotwash2} gives strong constraints on any new interactions. For chameleon models, these constraints are usually avoided due to the chameleon mechanism as long as typical test-masses used in the experiments have thin-shells.

Local gravity constraints for Model B Eq.~(\ref{modelb}) was found in \cite{Brax:2010kv}, see Fig.~(\ref{local_constraints}) for the constraints when $m=1$ and $n=6,10$.

In order for the chameleon to affect large scale structure formation we need the coupling, evaluated in the cosmological background $Q_{\phi 0} \equiv |\beta,_{\phi_{\rm today}}M_{\rm pl}|$, to be of gravitational strength: $\mathcal{O}(1)\lesssim Q_{\phi 0}$. 

For Model B, this corresponds to the r.h.s of the dashed line in Fig.~(\ref{local_constraints}). From this figure we see that this requires $M\lesssim 10^{-3}$ eV as claimed below Eq.~(\ref{modelb}).

Local gravity constraints for the original chameleon model (Model A with $m=1$) have been calculated in several papers, see e.g. \cite{Khoury:2003rn,Mota:2006fz,local1,local2,local3,local4,local5,MotaShaw}. 

Below, we derive thin-shell solutions for Model A Eq.~(\ref{modela}), and use these solutions to find the the local constraints coming from tests of the equivalence principle in the solar-system.
\subsubsection{Thin-shell solutions for the power-law coupling}
We show the existence of thin-shell solutions for the power-law coupling-function $\beta(\phi) = \left(\frac{\lambda\phi}{M_{\rm pl}}\right)^m$ together with an arbitrary run-away potential like e.g. $V(\phi) = \frac{M^{n+4}}{\phi^n}$ (Model A). This model was given a treatment in \cite{radion} and it was found that there do not exist thin-shell solutions for $m>1$ when $\lambda=\mathcal{O}(1)$. For values $\lambda \gg 1$, which is equivalent to a coupling scale $M_*=\frac{M_{\rm pl}}{\lambda} \ll M_{\rm pl}$, there can indeed exist thin-shells. Another conclusion the authors of \cite{radion} reached was the existence of a possible singularity in the field-profile. We have calculated the field-profile numerically and no such singularity was found. The solution corresponding to the claimed singularity seems to come from a mathematical correct, but non-physical, solution to the field equation.

The field equation in a static spherical symmetric metric with weak gravity reads
\begin{align}
	\frac{d^2\phi}{dr^2}+\frac{2}{r}\frac{d\phi}{dr} = V,_{\phi} + \rho\beta,_{\phi}
\end{align}
where we have assumed $\beta(\phi)\ll 1$. We consider a body with constant density $\rho_c$, radii $R_c$ and mass $M_c$ embedded in a background of homogeneous density $\rho_b$ and impose the boundary conditions
\begin{align}
\begin{array}{lcl}
	\left.\frac{d\phi}{dr}\right|_{r=0} &=& 0\\
	\left.\frac{d\phi}{dr}\right|_{r=\infty} &=& 0\\
	\phi(r\to\infty)&=&\phi_b
\end{array}
\end{align}
Lets start by considering a test body. The field inside and outside the body is then just a small perturbation in the background $\phi_b$. Solving the linearized field equations we find
\begin{align}
	\phi &= \phi_b - \frac{Q_b}{4\pi M_{\rm pl}}\frac{M_c}{r}e^{-m_br}~~~r>R_c\\
	Q_b &=  \beta,_{\phi_b}M_{\rm pl}\label{qb}
\end{align}
The amplitude of the fifth-force between two test bodies, located in the vacuum of space, is
\begin{align}
	F(r) = 2Q_b^2F_{\rm gravity}(r)~~~\text{for}~~~m_b r  < 1
\end{align}
Now lets see what happens for macroscopic bodies. The thin-shell solution for $m=1$ (see \cite{Khoury:2003rn}) is characterized by the field being stuck at the minimum, $\phi_c$, inside the body.  We therefore look for solutions where $\phi(0)\equiv \phi_i \approx \phi_c$. The linear approximation $V_{\rm eff,\phi} = m_c^2(\phi-\phi_c)$ is now valid close to $r=0$ with the solution
\begin{align}
	\phi &= \phi_c + \phi_c\frac{\sinh(m_c r)}{m_c r}\delta\\
	\delta &\equiv \frac{\phi_i-\phi_c}{\phi_c}
\end{align}
We assume that this solution is valid all the way to $r=R_c$. For this to be true, the linear term in the Taylor expansion of $V_{\rm eff, \phi}$ must dominate over the higher order terms inside the body. Since $\phi$ is increasing in $0<r<R_c$ the largest value of $|\phi-\phi_c|$ occurs at $r=R_c$, and leads to the condition
\begin{align}\label{cond}
	\frac{|\phi(R_c)-\phi_c|}{\phi_c} \ll \frac{2}{|n-m-3|}
\end{align}
Outside the body the linear approximation $V_{\rm eff,\phi}=m_b^2(\phi-\phi_b)$ is valid with the solution
\begin{align}
	\phi = \phi_b - \frac{AR_c}{r}e^{-m_b r}
\end{align}
where we have assumed $m_b R_c < 1$ as would be the case in most interesting cases. Matching the two solutions at $r=R_c$ gives us
\begin{align}
	A &= (\phi_b - \phi_c)\left(1+\frac{\tanh(m_c R_c)}{m_c R_c}\right)\\
	\delta &= \frac{\phi_b-\phi_c}{\phi_c\cosh(m_c R_c)}
\end{align}
The condition Eq.~(\ref{cond}) becomes
\begin{align}
	\frac{\phi_b-\phi_c}{\phi_c}\left(\frac{\tanh(m_c R_c)}{m_c R_c}\right) \ll \frac{2}{|n-m-3|}
\end{align}
For $x \gg 1$ we have $\tanh(x) \simeq 1$, and by using $\phi_b\gg \phi_c$ we find that the condition above is satisfied for all
\begin{align}
	m_c R_c \gg \frac{\phi_b}{\phi_c}
\end{align}
The far-away field can now be written
\begin{align}
	\phi = \phi_b - \frac{Q_{\rm eff}}{4\pi M_{\rm pl}}\frac{M_1}{r}e^{-m_br}
\end{align}
where
\begin{align}
	Q_{\rm eff} &= 3\beta,_{\phi_b}M_{\rm pl}\epsilon_{\rm th}\label{qeff}\\
	\epsilon_{\rm th} &= \frac{\phi_b-\phi_c}{6\beta,_{\phi_b}M_{\rm pl}^2\Phi_c}
\end{align}
and $\Phi_c$ is the gravitational potential for the body. The thin-shell factor $\epsilon_{\rm th}$ is on the same form as found in \cite{Khoury:2003rn}, but it does not have the geometrical interpretation as an explicit thin-shell. It is however this factor which determines the suppression of the fifth-force.

Comparing the effective coupling Eq.~(\ref{qeff}) (for 'large' bodies: $\epsilon_{\rm th}\ll 1$) with the corresponding expression Eq.~(\ref{qb}) (for 'small' bodies: $\epsilon_{\rm th}\gg 1$) we see that
\begin{align}
	\frac{Q_{\rm eff}}{Q_b}  = 3\epsilon_{\rm th}  \ll 1
\end{align}
This shows that the chameleon force, relative to gravity, between two bodies is suppressed as long as one (or both) of the bodies have a thin-shell, and demonstrates that the chameleon mechanism is present in this model. 


\subsubsection{Lunar Laser Ranging}
We will restrict our attention to tests of the equivalence principle using Lunar Laser Ranging (LLR), see e.g. \cite{Will:93}. LLR measures the free-fall acceleration of the moon and the earth relative to the sun. The acceleration induced by a fifth force with the field profile $\phi(r)$ and effective coupling $Q_{\rm eff }$ is $a_{\rm fifth} = |Q_{\rm eff}\nabla\phi |/M_{\rm pl}$.  In most interesting cases, $m_b^{-1}> 1 {\rm Au}$, the chameleon is a free field in the solar-system. This leads to the following constraint (see \cite{Khoury:2003rn} for a more detailed derivation)
\begin{align}
	\frac{2|a_{\rm moon}-a_{\oplus}|}{|a_{\rm moon}+a_{\oplus}|} \approx 2 Q_{\rm eff}^{\odot}|Q_{\rm eff}^{\rm m}-Q_{\rm eff}^{\oplus}| \lesssim 10^{-13}
\end{align}
where $Q_{\rm eff}^{\odot}$, $Q_{\rm eff}^{m}$ and $Q_{\rm eff}^{\oplus}$ is the effective coupling for the sun, moon and earth respectively. The background density in the solar-system is $\rho_b \approx 10^{-24}$ $g/cm^3$ corresponding to the average matter (dark and cold) density in our galaxy.

The resulting bounds for $m=2$ and $m=3$ are shown in Fig.~(\ref{fig:regimes}).\\


The strongest constraints on chameleon models (with natural parameters) typically comes from the Eot-Wash experiment \citep{eotwash}. As we can see in Fig.~(\ref{fig:regimes}), the interesting part of the parameter space is $n\ll 1$. In this regime the test-masses used in the Eot-Wash experiment will typically not have thin-shells and the experiment cannot, by design, detect the chameleon. Nevertheless, the LLR constraints are good enough to constraint this part of the parameter space where interesting cosmological signatures take place.

\begin{figure}%
\includegraphics[width=\columnwidth]{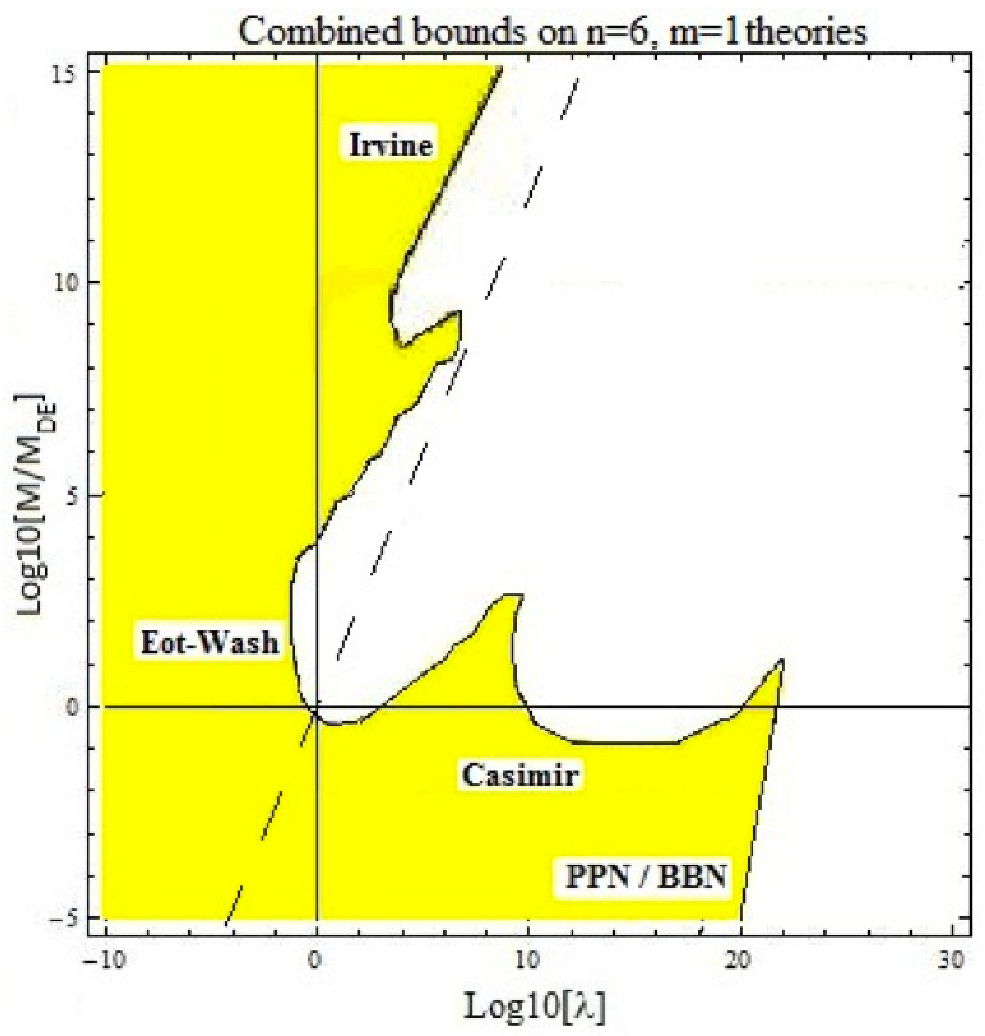}\\
\includegraphics[width=\columnwidth]{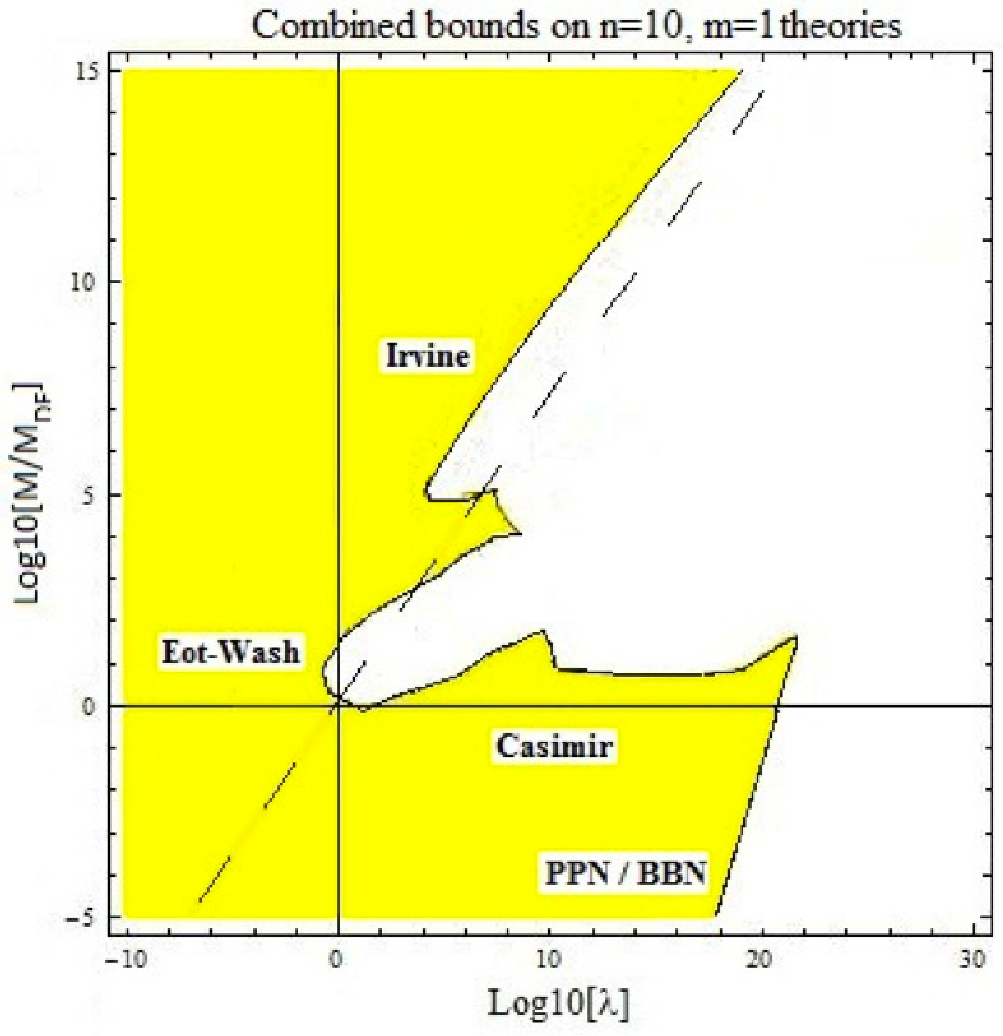}%
\caption{Local gravity constraints for the Model B Eq.~(\ref{modelb}) when $n=6$ (top) and $n=10$ (bottom). The horizontal line shows $M=M_{\rm DE}=10^{-3}$eV. The vertical line shows $\lambda=1$. The dashed line shows $Q_{\phi 0} = |\beta,_{\phi_{\rm today}}|M_{\rm pl}=1$.  The labels 'Irvine' and 'Eot-Wash' refers to test of the Newtonian gravitational law, 'Casimir' refers to Casimir experiments, 'PPN' refers to test of post-Newtonian gravity in the solar-system and 'BBN' to constraints from Big Bang Nucleosynthesis.}%
\label{local_constraints}%
\end{figure}


\subsection{Cosmological constraints}
Due to the conformal coupling, Eq.~(\ref{conformal_coupling}), of $\phi$ to matter, a constant mass scale $m_0$ in the Jordan-frame is related to a $\phi$-dependent mass scale $m(\phi)$ in Einstein-frame by $m(\phi)=m_0 e^{\beta(\phi)}$. A variation in $\phi$ leads to a variation in the various masses
\begin{align}
	\left|\frac{\Delta m}{m}\right| \approx \Delta\beta(\phi)
\end{align}
BBN constrains the variation in $m(\phi)$ from the time of nucleosynthesis until today to be less than around $10\%$.

If the field has settled at the minimum before BBN, the resulting bound turns into $\beta(\phi_{\rm today})\lesssim 0.1$ since $\beta(\phi)$ is an increasing function of time. When the field is not at the minimum at BBN we must also require $\beta(\phi_{\rm BBN})\lesssim 0.1$. 

For our models this constraints only the parameters where the coupling satisfies $Q_{\phi 0} = |\beta,_{\phi_{\rm today}}M_{\rm pl}| \gg 1$, in which both the background and the matter perturbations are completely similar to $\Lambda\rm CDM$.  The bounds for Model B Eq.~(\ref{modelb}) are shown in Fig.~(\ref{local_constraints}).

Another important restriction on chameleon theories comes out from considering the isotropy of the Cosmological Microwave Background (CMB) \citep{cmb_49}. A difference in the value of $\phi$ today and the value it had during the epoch of recombination would mean that the electron mass at that epoch differed from its present value by $\frac{\Delta m_e}{m_e} \approx \Delta\beta(\phi)$. Such a change in $m_e$ would, in turn, alter the redshift at which recombination occurred, $z_{\rm rec}$:
\begin{align}
	\left|\frac{\Delta z_{\rm rec}}{z_{\rm rec}}\right| \approx \Delta \beta(\phi)
\end{align}
WMAP bounds $z_{\rm rec}$ to be within $5\%$ (at $2\sigma$, 23\% at $4\sigma$) of the value that has been calculated using the present day value of $m_e$ \citep{wmapbound}. Denoting $\phi_0$, $\phi_{\rm rec}$ and $\phi_{\rm BBN}$ with the field value today, at recombination and BBN respectively. Then $\beta(\phi_0) > \beta(\phi_{\rm rec}) > \beta(\phi_{\rm BBN})$ and the CMB bound is weaker then the bound coming from BBN.


\section{Perturbations}
In this section we will study the growth of perturbations in chameleon models. We start by consider the general scalar-tensor model given by the action Eq.~(\ref{cham_action_1}) with a universal matter coupling-function $\beta(\phi)$ and potential $V(\phi)$. In deriving the perturbation we will work in units of $M_{\rm pl}= \frac{1}{\sqrt{8\pi G}}\equiv 1$. For simplicity, we will consider the Jordan-frame matter-density satisfying
\begin{equation}
\dot{\rho}_m + \left(3H-\beta,_{\phi}\dot{\phi}\right)\rho_m=0
\end{equation}
since this choice will simplify the field equation. In terms of the Einstein-frame density $\rho_m^{EF}$ this choice corresponds to $\rho_m=e^{\beta(\phi)}\rho_m^{EF}$. This is just a matter of convenience since $e^{\beta(\phi)}\approx 1$ in the late universe whenever the theory satisfies the BBN bounds. With this choice the field equation reads
\begin{equation}
\ddot{\phi}+3H\dot{\phi} + V,_{\phi} + \beta,_{\phi}\rho_m=0
\end{equation}
The most general metric in a FLRW space time with scalar perturbations is given by
\begin{align}
ds^2=-(1+2\alpha){dt}^2-2 a B_{,i}{dt}{dx}^i\\
+a^2\left((1+2\psi)\delta_{ij}+2\gamma_{,i;j}\right){d x}^i{d x}^j\nonumber
\end{align}
where the covariant derivative is given in terms of the three-space metric which in the case of a flat background reduces to $\delta_{ij}$. We decompose the field $\phi$ into the background and perturbations parts: $\phi({\bf x},t) = \overline{\phi}(t) +\delta\phi({\bf x},t)$. The EM-tensors of non-relativistic matter can be decomposed as
\begin{align}
T^0_0 = -\rho_m(1+\delta_m), ~~~T^0_i = -\rho_m v,_i
\end{align}
where $v$ is the peculiar velocity of non-relativistic matter and $\delta_m$ is the matter-density perturbations defined by
\begin{equation}
\delta_m\equiv \frac{\delta \rho_m}{\rho_m}-\frac{\dot{\rho}_m}{\rho_m}v\equiv \frac{\delta\rho_m}{\rho_m}~~~\text{in the co-moving gauge}
\end{equation}
The equation determining the evolution of the perturbations follows from the Einstein-equations. In the gauge-ready formulation \citep{Hwang:1991aj}, the scalar perturbations equations are\\
\begin{widetext}
\begin{align}
\label{p1}\ddot{\delta\phi}+3H\dot{\delta\phi}+(V_{,\phi\phi}-\frac{\Delta}{a^2})\delta\phi+\beta,_{\phi\phi}\rho_m\delta\phi+2\alpha V_{,\phi}+\beta,_{\phi}(2\alpha\rho_m+\delta\rho_m)-\dot{\phi}(\dot{\alpha}-3H\alpha+\kappa)&=&0\\
\label{p2}\dot{\delta \rho_m}+3 H \delta\rho_m-\rho_m \left(\kappa-3H\alpha+\frac{\Delta}{a^2}v\right)- \beta,_{\phi}(\rho_m\dot{\delta\phi}+\delta\rho_m\dot{\phi})-\beta,_{\phi\phi}\rho_m\dot{\phi}\delta\phi &=&0\\
\label{p3}\dot{\kappa}+2H\kappa+3\alpha\dot{H}+\frac{\Delta}{a^2}\alpha-\frac{1}{2}\left(\delta\rho_m-4\alpha\dot{\phi}^2+4\dot{\phi}\dot{\delta \phi}-2V_{,\phi}\delta\phi\right)&=&0\\
\label{p4}H\kappa+\frac{\Delta}{a^2}\psi-\frac{1}{2}\left(-\delta \rho_m+\alpha \dot{\phi}^2-\dot{\phi}\dot{\delta\phi}-V_{,\phi}\delta\phi\right)&=&0\\
\label{p5}\kappa+\frac{\Delta}{a^2}\chi-\frac{3}{2}(\rho_m v+\dot{\phi}\delta\phi)&=&0\\
\label{p6}\dot{v}-\alpha+\beta,_{\phi}(\dot{\phi}v-\delta\phi)&=&0\\
\label{p7}\dot{\chi}+H\chi-\alpha-\psi&=&0
\end{align}
\end{widetext}
with
\begin{align}
\chi&=a(B+a\dot{\gamma})\\
\kappa&=3(-\dot{\psi}+H\alpha)-\frac{\Delta}{a^2}\chi
\end{align}
and $\Delta$ being the co-moving covariant three-space Laplacian. In the list of equations above Eq.~(\ref{p1}) is the scalar field equation of motion, Eq.~(\ref{p2}) the continuity equation, Eq.~(\ref{p3}) the Raychauhuri equation, Eq.~(\ref{p4}) the ADM energy constraint, Eq.~(\ref{p5}) the momentum conservation constraint and Eq.~(\ref{p7}) the ADM propagation equation. In these equations we have not yet fixed the gauge-degrees of freedom. The choice of a gauge will simplify the system and we will work in the so-called co-moving gauge $(v=0)$. This gauge leaves no residual gauge freedom and we can solve the system for the two variables $(\delta \phi,\delta_m)$ directly. 

From Eq.~(\ref{p6}) we have $\alpha = -\beta,_{\phi}\delta\phi$. Solving Eq.~(\ref{p3}) for $\kappa$ and inserting this into Eq.~(\ref{p1},\ref{p3}) we find, after transforming to Fourier space, the following equations\\
\begin{widetext}
\begin{align}\label{fulldmeq}
\ddot{\delta}_m+&2H\dot{\delta}_m-\frac{1}{2}\rho_m\delta_m+\delta\phi\left(V_{,\phi}-\beta,_{\phi}[6H^2+6\dot{H}-\frac{k^2}{a^2}+2\dot{\phi}^2]\right)\nonumber\\
&-\delta{\phi}\left(\beta,_{\phi\phi}[2H\dot{\phi}-V_{\rm eff,\phi}]+\beta,_{\phi\phi\phi}\dot{\phi}^2\right)-\beta,_{\phi}\ddot{\delta\phi}-\dot{\delta\phi}\left(5\beta,_{\phi}H+2\dot{\phi}+2\beta,_{\phi\phi}\dot{\phi}\right)=0\
\end{align}
\begin{align}\label{full_pert_eq}
&\ddot{\delta\phi}+(3H+2\beta,_{\phi}\dot{\phi})\dot{\delta \phi}+\beta,_{\phi}\rho_m\delta_m-\dot{\phi}\dot{\delta}_m+\left(m_{\phi}^2+\frac{k^2}{a^2}-2\beta,_{\phi}V_{\rm eff,\phi}+2\beta,_{\phi\phi}\dot{\phi}^2\right)\delta\phi=0
\end{align}
\end{widetext}
where $k$ is a co-moving wavenumber.

When the field is slow rolling along the minimum we can neglect all terms proportional to $\dot{\phi}$ and the oscillating term $V_{\rm eff,\phi}$. The perturbations in $\phi$ will evolve more slowly than the perturbations in $\delta_m$ for scales deep inside the Hubble radius, thus, the term $\rho_m\beta,_{\phi}\delta_m$ and $(m_{\phi}^2+\frac{k^2}{a^2})\delta\phi$ will dominate over the $\delta\phi$ time derivatives in Eq.~(\ref{full_pert_eq}). Using these approximations, we can simplify Eq.~(\ref{fulldmeq}) and Eq.~(\ref{full_pert_eq}) to
\begin{align}\label{newtonian_pert_approx}
\begin{array}{l}
\ddot{\delta_m} + 2H\dot{\delta_m} = \frac{3}{2}\Omega_m H^2\left(1+\frac{2\beta,_{\phi}^2}{1+\frac{a^2m_{\phi}^2}{k^2}}\right)\delta_m\\
\delta\phi = -3\beta,_{\phi}\Omega_m\left(\frac{H^2}{m_{\phi}^2}\right) \frac{1}{1 + \frac{k^2}{a^2m_{\phi}^2}}
\end{array}
\end{align}
Note that the perturbations in $\phi$ satisfies $\delta\phi \ll \delta_m$ (in $M_{\rm pl}=1$ units) as long as $m_{\phi}^2\gg H^2$ and $\beta,_{\phi}\lesssim \mathcal{O}(1)$ justifying dropping the $\delta\phi$-derivatives in Eq.~(\ref{full_pert_eq}).

Restoring $M_{\rm pl}^{-2}\equiv 8\pi G$ and defining $Q_{\phi} = |\beta,_{\phi}M_{\rm pl}|$ we can write this first equation on the same form as in $\Lambda\rm CDM$
\begin{align}
\ddot{\delta_m} + 2H\dot{\delta_m} = 4\pi G_{\text{eff}}\rho_m\delta_m
\end{align}
where
\begin{align}\label{geff}
G_{\rm eff} = G\left(1+\frac{2Q_{\phi}^2}{1+\frac{a^2m_{\phi}^2}{k^2}}\right)
\end{align}
The quantity $G_{\rm eff}$ is seen to encode the modification of gravity due to the chameleon in the weak-field regime.

The chameleon will also exhibit an oscillating term, but this term is time-decreasing and hence negligible for small redshifts. In some $f(R)$-models however, this oscillating term can grow to infinity because the mass of the scalaron is not bounded above. The divergence of this mass can be removed by adding a UV-term as shown in \cite{Thongkool:2009js}.

In studying perturbations, it is convenient to introduce the growth-factor $f = \frac{d\log(\delta_m)}{d\log(a)}$. In $\Lambda$CDM $f\to 1$ at high redshifts and $f\to 1$ in an Einstein-de Sitter universe. It is important to find a characteristics in the perturbations that can discriminate between different DE models and the $\Lambda$CDM. Writing the growth factor as
\begin{align}
f = \Omega_m^{\gamma}
\end{align}
can be a parametrization that is useful for this purpose, see e.g. \cite{Gannouji:2009tx,Gannouji:2008wt}. In $\Lambda$CDM we have to a good accuracy $\gamma\approx 0.55$ for low redshifts. Of course in some models $\gamma$ will vary to much for it to be considered a constant, and there can also be a scale dependence, so we should write $\gamma = \gamma(z,k)$.

We will be most interested in scales $k$ relevant to the galaxy power spectrum \citep{Tsujikawa:2009ku}
\begin{align}
0.01 h Mpc^{-1}\lesssim k \lesssim 0.2 h Mpc^{-1}
\end{align}
where $h=0.72\pm 0.08$ corresponds to the uncertainty in the Hubble factor today. These scales are also in the linear regime of perturbations.

It is also convenient to introduce the length scale of the perturbations $\lambda_p = \frac{2\pi a}{k}$ and the length scale of the chameleon $\lambda_{\phi} = \frac{2\pi}{m_{\phi}}$. In Eq.~(\ref{geff}) we have two asymptotic regimes:
\begin{align}
G_{\rm eff} = \left\{\begin{array}{lr}G & \lambda_p \gg \lambda_{\phi}\\G(1+2Q_{\phi}^2) &  \lambda_p \ll \lambda_{\phi}\end{array}\right.
\end{align}
In the GR regime, $ \lambda_p \gg \lambda_{\phi}$, the perturbations show no deviation from $\Lambda\rm CDM$.

In the scalar regime, $ \lambda_p \ll \lambda_{\phi}$, however, the matter-perturbations will feel a stronger gravitational constant than in GR. The coupling $Q_{\phi}$ is in general a dynamical quantity, which will increase with time when the chameleon follows the minimum. Thus, when we reach a time where $Q_{\phi}> 1$ the perturbations will start to grow with increasing amplitude and will quickly enter the non-linear regime.

See \cite{baldi,Li1,Li2,Li3,Li4,Li5,vale1} for a numerical N-body simulation analysis with this type of models, and \cite{nunes,vale,void,phil} for a study of the spherical collapse in cosmological models with a time dependent coupling between dark energy, dark matter and other matter fields.


\subsection{The critical length scale $\lambda_{\phi}$}
In order to study the perturbations more closely, we look at the value of the critical length scale today $\lambda_{\phi,0}$ for our models.

\begin{figure}%
\centering
\includegraphics[width=\columnwidth]{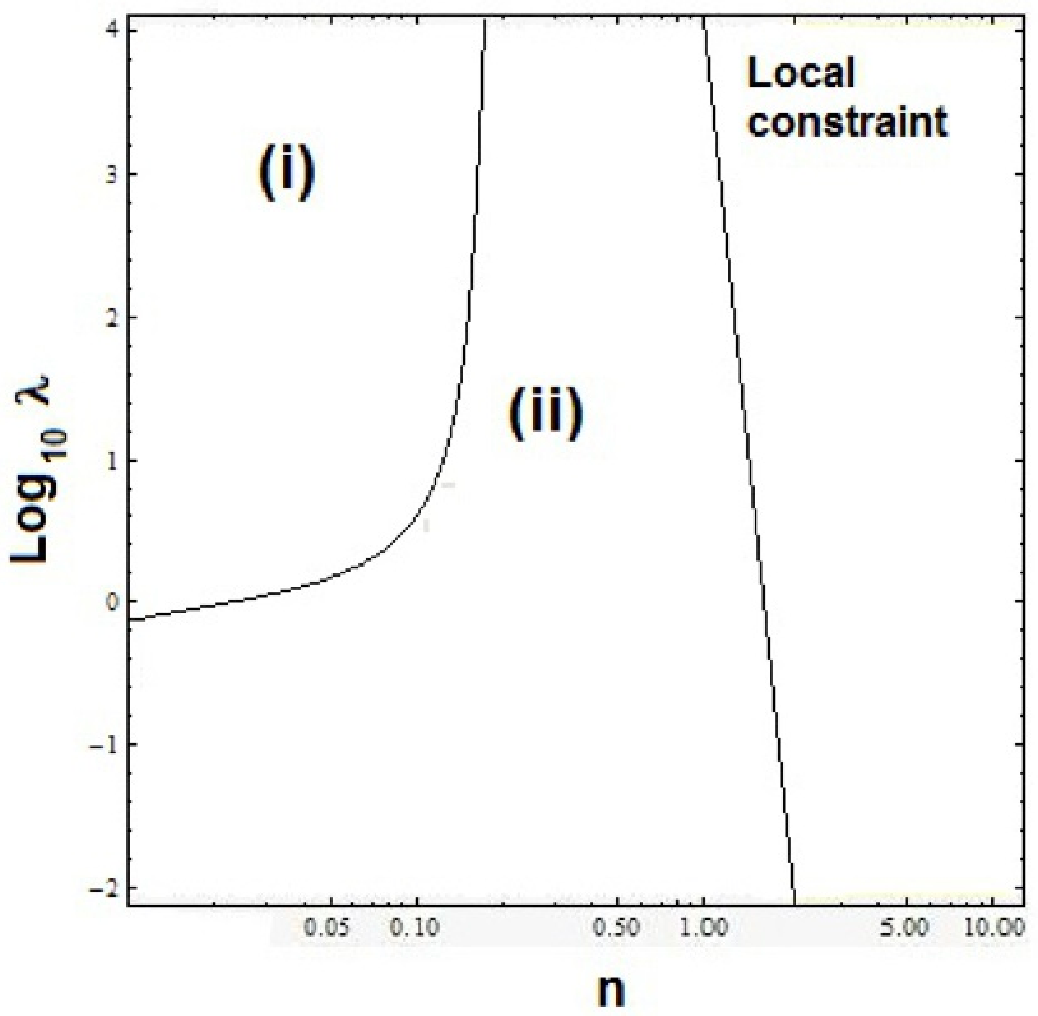}\\
\includegraphics[width=\columnwidth]{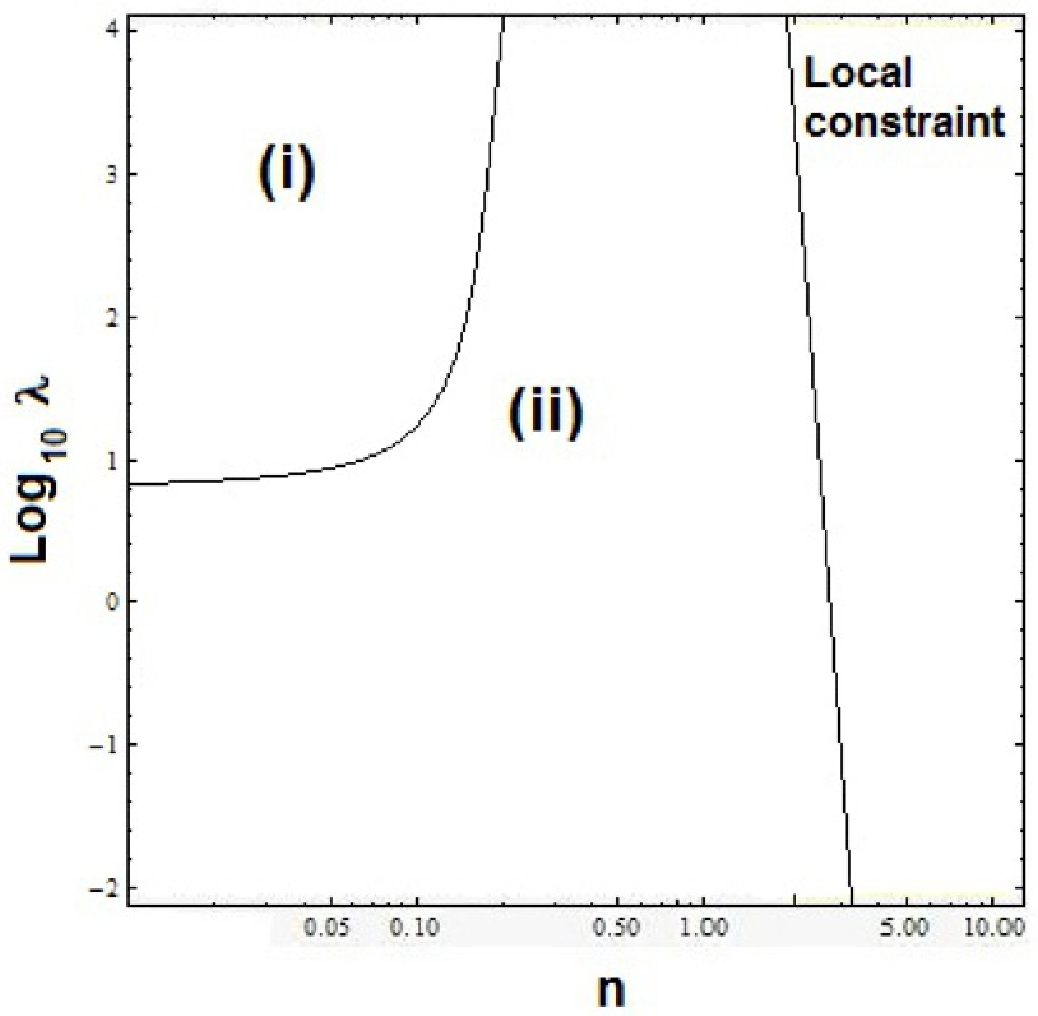}%
\caption{The two regimes for the growth factor $\gamma_0$ in Model A Eq.~(\ref{modela}): (i) $\gamma < 0.50$, and (ii) $\gamma \approx 0.55$ for all relevant scales. In region $(i)$ we have a significant dispersion between scales. Above, the quadratic coupling $m=2$, and below the cubic coupling $m=3$. The label 'Local constraints' shows the allowed region from experimental tests of the equivalence principle (Lunar Laser Ranging). Due to numerical difficulties the perturbations have been integrated using the approximation Eq.~(\ref{newtonian_pert_approx}) rather than using the full equations Eq.~(\ref{fulldmeq})-(\ref{full_pert_eq})}%
\label{fig:regimes}%
\end{figure}

The critical length scale for Model A Eq.~(\ref{modela}) satisfies
\begin{align}
	\lambda_{\phi,0} \sim 10^{-5 +\frac{15}{n+1}}|Q_{\phi 0}|^{-\frac{(n+2)}{2(n+1)}}{\rm pc}
\end{align}
where $Q_{\phi 0} = |\beta,_{\phi_{\rm today}}M_{\rm pl}|$. In order for this length scale to affect the matter perturbations: $\lambda_{\phi,0} = \mathcal{O}(1{\rm Mpc})$ and $Q_{\phi 0}\sim 1$, we need to impose $n<0.5$. See Fig.~\ref{fig:regimes} for a plot of the growth factor $\gamma(z=0)$. The plot shows the two regimes:

\begin{itemize}
	\item (i): Phase space where $\gamma < 0.50$ for all relevant scales. The perturbations are in the scalar regime.
	\item (ii): Phase space where $\gamma \approx 0.55$ for all relevant scales. This is the GR regime.
\end{itemize}

For our Model B Eq.~(\ref{modelb}), the critical length scale is given by
\begin{align}
	\lambda_{\phi,0} = \sqrt{\frac{m}{n(n+m)}} \left(\frac{M_{\rm pl}}{M}\right)^{\frac{(n-2)}{2(n-1)}} Q_{\phi 0}^{-\frac{(n-2)}{2(n-1)}} \frac{2\pi}{M}
\end{align}
which gives
\begin{align}
	\lambda_{\phi,0} \sim 10^{-5 -\frac{15}{n-1}}Q_{\phi 0}^{-\frac{(n-2)}{2(n-1)}}{\rm pc}
\end{align}
In order to have $\lambda_{\phi,0} = \mathcal{O}(1{\rm Mpc})$ together with a coupling $Q_{\phi 0}$ of the order of unity we need $n\lesssim 0.5$. However, for $n<2$ the model is no longer a chameleon according to our definition in section 1: the range of the field is shorter in the low density cosmological background than in a high density environment. This also means that local gravity bounds will most certainly be violated.

The only way to increase $\lambda_{\phi,0}$ up to a mega-parsec value is by decreasing the coupling strength. For $n>2$ we need 
\begin{align}
	Q_{\phi 0} \lesssim 10^{-10}
\end{align}
which is to small to significantly affect the growth of the perturbations. Thus the perturbations in Model B are always in the GR regime.

It is only in Model A Eq.~(\ref{modela}) that we can have interesting signatures on the matter perturbations. But after imposing local gravity constraints we find that the perturbations are confined to be in the GR regime with no signature on the matter perturbations or on the background evolution relative to $\Lambda \rm CDM$. This agrees with the result found in \cite{2010arXiv1010.3769G} (for $m=1$). The only way to have observable signatures in these models is to restrict the coupling to dark matter only, and thereby avoiding the local constraints.

The different regimes shown in Fig.~(\ref{fig:regimes}) have been derived by considering an universal coupling. Since dark matter is dominating over baryonic matter at large scales the regimes in this figure is expected to be similar if we restrict the coupling to dark matter only.

In \cite{2010arXiv1010.3769G}, the chameleon model
\begin{align}
	\beta(\phi) &= \frac{\lambda\phi}{M_{\rm pl}}\\
	V(\phi) &= M^4(1-\mu(1-e^{-\frac{\phi}{M_{\rm pl}}})^n)
\end{align}
where $0<\mu<1$ and $0<n<1$, was found to have observable signatures on the growth of the matter perturbations even when local gravity constraints was taken into account. We note that this potential do not directly generalize to a more general coupling-function $\beta(\phi) = \left(\frac{\lambda\phi}{M_{\rm pl}}\right)^m$. This is because the requirement $0<n<1$, which is required in order to have a positive definite mass of the field, leads to violation of local gravity constraints for parameters where signatures are present.

\section{Summary and Conclusions}
We have discussed the cosmological evolution of chameleon models with power-law couplings and power-law potentials. The chameleon follows the attractor solution $\phi=\phi_{\rm min}$ as long as $m_{\phi} \gg H$, and the attractor can be reached for a large span of initial conditions. In fact, non-constant couplings can allow for a larger off-set from the attractor in the early universe and still be in agreement with BBN bounds. Along the attractor the chameleon is slow rolling and can account for the late time acceleration of the universe. The background evolution is however found to be very close to that of $\Lambda\rm CDM$, as found in many other similar models (see e.g. \cite{de2,Brax:2008hh}). The mass-scale $M$ in the potential is fine-tuned in the same manner as a cosmological constant, thus not providing a solution to the fine-tuning problem. This is however motivated and required by local gravity experiments.

Even though the background expansion is very close $\Lambda \rm CDM$, the growth of the linear perturbations can be quite different. The reason is the fifth-force acting on both dark matter and baryons, which leads to a different growth rate of matter perturbations on cosmic scales.

For our Model B Eq.~(\ref{modelb}), the linear perturbations are not affected by the chameleon since the field-range is in general too small compared to cosmic scales, or, if the field-range is large enough then the coupling is too small to produce observable effects. Otherwise the model would be ruled out by local gravity constraints.

In Model A Eq.~(\ref{modela}), the range of the chameleon can be large enough as to affect the matter perturbations, which leads to a growth-rate different from $\Lambda \rm CDM$. Since the coupling, in general, varies with time, we will also have a dispersion for scales within the linear regime. However, for this to be the case, it must be emphasized that local gravity constraints force us to have a gravitational coupling of the chameleon field to dark matter only. By neglecting the coupling to baryonic matter, the purpose of the chameleon mechanism are lost. But with this consideration, the growth of matter perturbations can in principle allow us to discriminate between Model A and $\Lambda\rm CDM$. 

Naively, one would expect that introducing a more general coupling would lead to a more richer phenomenology. The reason why this is not the case in the models considered here is because local constraints are so stringent it forces the chameleon to stay at the local minimum almost everywhere in space. This again means that the kinetic term of the chameleon is suppressed and the dynamics of the theory is determined solely by the interplay between the potential and the coupling. In the power-law models considered here this interplay is very similar to the standard chameleon model \cite{Brax:2010kv}.

If future galaxy surveys manages to pin down the matter power spectrum to a greater accuracy, and detect a clear deviation from $\Lambda\rm CDM$ (especially if a scale dependence in the growth index $\gamma$ is detected), then it would be interesting to see how good these chameleon models can fit the data.

Even though these models, for most parameters choices, do not leave an imprint on the linear matter perturbations, it may have an impact on the small scale structure formation. It would be interesting to investigate the effect of these models on the non-linear regime of structure formation by means of high resolution N-body simulations. We leave this for future work.


\section{Acknowledgments}
We thank P. Brax, C. van de Bruck and N. J. Nunes for useful comments and discussions. DFM thanks the Research Council of Norway FRINAT grant 197251/V30. HAW thanks the Research Council of Norway FRINAT grant 197251/V30.\\


\bibliography{bibfile}

\begin{thebibliography}{66}
\expandafter\ifx\csname natexlab\endcsname\relax\def\natexlab#1{#1}\fi

\bibitem[{{Baldi}(2010)}]{baldi}
{Baldi}, M. 2010, astro-ph.CO/1005.2188

\bibitem[{Baldi {et~al.}(2010)Baldi, Pettorino, Robbers, \& Springel}]{vale1}
Baldi, M., Pettorino, V., Robbers, G., \& Springel, V. 2010, Mon. Not. Roy.
  Astron. Soc., 403, 1684

\bibitem[{{Bean} {et~al.}(2007){Bean}, {Bernat}, {Pogosian}, {Silvestri}, \&
  {Trodden}}]{de4}
{Bean}, R., {Bernat}, D., {Pogosian}, L., {Silvestri}, A., \& {Trodden}, M.
  2007, \prd, 75, 064020

\bibitem[{Bourliot {et~al.}(2007)Bourliot, Ferreira, Mota, \& Skordis}]{de12}
Bourliot, F., Ferreira, P.~G., Mota, D.~F., \& Skordis, C. 2007, Phys. Rev.,
  D75, 063508

\bibitem[{Brax {et~al.}(2010)Brax, Rosenfeld, \& Steer}]{phil}
Brax, P., Rosenfeld, R., \& Steer, D.~A. 2010, JCAP, 1008, 033

\bibitem[{{Brax} {et~al.}(2004{\natexlab{a}}){Brax}, {van de Bruck}, \&
  {Davis}}]{radion}
{Brax}, P., {van de Bruck}, C., \& {Davis}, A. 2004{\natexlab{a}}, JCAP, 11, 4

\bibitem[{{Brax} {et~al.}(2004{\natexlab{b}}){Brax}, {van de Bruck}, {Davis},
  {Khoury}, \& {Weltman}}]{Brax:2004qh}
{Brax}, P., {van de Bruck}, C., {Davis}, A., {Khoury}, J., \& {Weltman}, A.
  2004{\natexlab{b}}, \prd, 70, 123518

\bibitem[{{Brax} {et~al.}(2007){Brax}, {van de Bruck}, {Davis}, {Mota}, \&
  {Shaw}}]{local3}
{Brax}, P., {van de Bruck}, C., {Davis}, A., {Mota}, D.~F., \& {Shaw}, D. 2007,
  \prd, 76, 124034

\bibitem[{{Brax} {et~al.}(2009){Brax}, {van de Bruck}, {Davis}, \&
  {Shaw}}]{local1}
{Brax}, P., {van de Bruck}, C., {Davis}, A., \& {Shaw}, D. 2009,
  hep-ph/0911.1086

\bibitem[{{Brax} {et~al.}(2008){Brax}, {van de Bruck}, {Davis}, \&
  {Shaw}}]{Brax:2008hh}
{Brax}, P., {van de Bruck}, C., {Davis}, A., \& {Shaw}, D.~J. 2008, \prd, 78,
  104021

\bibitem[{Brax {et~al.}(2007)Brax, van~de Bruck, Davis, Mota, \& Shaw}]{local4}
Brax, P., van~de Bruck, C., Davis, A.-C., Mota, D.~F., \& Shaw, D.~J. 2007,
  Phys. Rev., D76, 085010

\bibitem[{{Brax} {et~al.}(2010{\natexlab{a}}){Brax}, {van de Bruck}, {Davis},
  {Shaw}, \& {Iannuzzi}}]{Brax:2010xx}
{Brax}, P., {van de Bruck}, C., {Davis}, A.~C., {Shaw}, D.~J., \& {Iannuzzi},
  D. 2010{\natexlab{a}}, Physical Review Letters, 104, 241101

\bibitem[{{Brax} {et~al.}(2010{\natexlab{b}}){Brax}, {van de Bruck}, {Mota},
  {Nunes}, \& {Winther}}]{Brax:2010kv}
{Brax}, P., {van de Bruck}, C., {Mota}, D.~F., {Nunes}, N.~J., \& {Winther},
  H.~A. 2010{\natexlab{b}}, astro-ph.CO/1006.2796

\bibitem[{{Brax} \& {Zioutas}(2010)}]{solarcham}
{Brax}, P., \& {Zioutas}, K. 2010, \prd, 82, 043007

\bibitem[{{Carroll} {et~al.}(2006){Carroll}, {Sawicki}, {Silvestri}, \&
  {Trodden}}]{de1}
{Carroll}, S.~M., {Sawicki}, I., {Silvestri}, A., \& {Trodden}, M. 2006, New
  Journal of Physics, 8, 323

\bibitem[{{Copeland} {et~al.}(2006){Copeland}, {Sami}, \&
  {Tsujikawa}}]{copeland}
{Copeland}, E.~J., {Sami}, M., \& {Tsujikawa}, S. 2006, International Journal
  of Modern Physics D, 15, 1753

\bibitem[{{Davis} {et~al.}(2009){Davis}, {Schelpe}, \& {Shaw}}]{cp1}
{Davis}, A., {Schelpe}, C.~A.~O., \& {Shaw}, D.~J. 2009, \prd, 80, 064016

\bibitem[{{Davis} {et~al.}(2010){Davis}, {Schelpe}, \& {Shaw}}]{cp3}
---. 2010, ArXiv e-prints

\bibitem[{{Davis} {et~al.}(2007){Davis}, {M{\"o}rtsell}, {Sollerman}, {Becker},
  {Blondin}, {Challis}, {Clocchiatti}, {Filippenko}, {Foley}, {Garnavich},
  {Jha}, {Krisciunas}, {Kirshner}, {Leibundgut}, {Li}, {Matheson}, {Miknaitis},
  {Pignata}, {Rest}, {Riess}, {Schmidt}, {Smith}, {Spyromilio}, {Stubbs},
  {Suntzeff}, {Tonry}, {Wood-Vasey}, \& {Zenteno}}]{wood}
{Davis}, T.~M., {et~al.} 2007, \apj, 666, 716

\bibitem[{{Durrer} \& {Maartens}(2008)}]{durrer}
{Durrer}, R., \& {Maartens}, R. 2008, General Relativity and Gravitation, 40,
  301

\bibitem[{{Faulkner} {et~al.}(2007){Faulkner}, {Tegmark}, {Bunn}, \&
  {Mao}}]{de2}
{Faulkner}, T., {Tegmark}, M., {Bunn}, E.~F., \& {Mao}, Y. 2007, \prd, 76,
  063505

\bibitem[{{Gannouji} {et~al.}(2010){Gannouji}, {Moraes}, {Mota}, {Polarski},
  {Tsujikawa}, \& {Winther}}]{2010arXiv1010.3769G}
{Gannouji}, R., {Moraes}, B., {Mota}, D.~F., {Polarski}, D., {Tsujikawa}, S.,
  \& {Winther}, H.~A. 2010, ArXiv e-prints

\bibitem[{{Gannouji} {et~al.}(2009{\natexlab{a}}){Gannouji}, {Moraes}, \&
  {Polarski}}]{Gannouji:2009tx}
{Gannouji}, R., {Moraes}, B., \& {Polarski}, D. 2009{\natexlab{a}}, ArXiv
  e-prints

\bibitem[{{Gannouji} {et~al.}(2009{\natexlab{b}}){Gannouji}, {Moraes}, \&
  {Polarski}}]{Gannouji:2008wt}
---. 2009{\natexlab{b}}, JCAP, 2, 34

\bibitem[{Gies {et~al.}(2008)Gies, Mota, \& Shaw}]{local5}
Gies, H., Mota, D.~F., \& Shaw, D.~J. 2008, Phys. Rev., D77, 025016

\bibitem[{{Hinshaw} {et~al.}(2003){Hinshaw}, {Spergel}, {Verde}, {Hill},
  {Meyer}, {Barnes}, {Bennett}, {Halpern}, {Jarosik}, {Kogut}, {Komatsu},
  {Limon}, {Page}, {Tucker}, {Weiland}, {Wollack}, \& {Wright}}]{cmb_49}
{Hinshaw}, G., {et~al.} 2003, \apjs, 148, 135

\bibitem[{{Hoyle} {et~al.}(2001){Hoyle}, {Schmidt}, {Heckel}, {Adelberger},
  {Gundlach}, {Kapner}, \& {Swanson}}]{eotwash2}
{Hoyle}, C.~D., {Schmidt}, U., {Heckel}, B.~R., {Adelberger}, E.~G.,
  {Gundlach}, J.~H., {Kapner}, D.~J., \& {Swanson}, H.~E. 2001, Physical Review
  Letters, 86, 1418

\bibitem[{{Huterer} \& {Linder}(2007)}]{linder2}
{Huterer}, D., \& {Linder}, E.~V. 2007, \prd, 75, 023519

\bibitem[{{Hwang}(1991)}]{Hwang:1991aj}
{Hwang}, J. 1991, \apj, 375, 443

\bibitem[{{Kapner} {et~al.}(2007){Kapner}, {Cook}, {Adelberger}, {Gundlach},
  {Heckel}, {Hoyle}, \& {Swanson}}]{eotwash}
{Kapner}, D.~J., {Cook}, T.~S., {Adelberger}, E.~G., {Gundlach}, J.~H.,
  {Heckel}, B.~R., {Hoyle}, C.~D., \& {Swanson}, H.~E. 2007, Physical Review
  Letters, 98, 021101

\bibitem[{{Khoury} \& {Weltman}(2004)}]{Khoury:2003rn}
{Khoury}, J., \& {Weltman}, A. 2004, \prd, 69, 044026

\bibitem[{Koivisto \& Mota(2008)}]{de11}
Koivisto, T., \& Mota, D.~F. 2008, JCAP, 0806, 018

\bibitem[{{Koyama} {et~al.}(2009){Koyama}, {Taruya}, \& {Hiramatsu}}]{de9}
{Koyama}, K., {Taruya}, A., \& {Hiramatsu}, T. 2009, \prd, 79, 123512

\bibitem[{{Lahav} {et~al.}(1991){Lahav}, {Lilje}, {Primack}, \& {Rees}}]{a2}
{Lahav}, O., {Lilje}, P.~B., {Primack}, J.~R., \& {Rees}, M.~J. 1991, \mnras,
  251, 128

\bibitem[{Li \& Barrow(2010)}]{Li3}
Li, B., \& Barrow, J.~D. 2010, 1005.4231 [astro-ph.CO]

\bibitem[{Li {et~al.}(2010{\natexlab{a}})Li, Mota, \& Barrow}]{Li1}
Li, B., Mota, D.~F., \& Barrow, J.~D. 2010{\natexlab{a}}, 1009.1400
  [astro-ph.CO]

\bibitem[{Li {et~al.}(2010{\natexlab{b}})Li, Mota, \& Barrow}]{Li2}
---. 2010{\natexlab{b}}, 1009.1396 [astro-ph.CO]

\bibitem[{Li \& Zhao(2009)}]{Li4}
Li, B., \& Zhao, H. 2009, Phys. Rev., D80, 044027

\bibitem[{{Linder}(2005)}]{linder1}
{Linder}, E.~V. 2005, \prd, 72, 043529

\bibitem[{Mota(2008)}]{de13}
Mota, D.~F. 2008, JCAP, 0809, 006

\bibitem[{Mota \& Barrow(2004)}]{barrow}
Mota, D.~F., \& Barrow, J.~D. 2004, Phys. Lett., B581, 141

\bibitem[{Mota {et~al.}(2007)Mota, Kristiansen, Koivisto, \& Groeneboom}]{de10}
Mota, D.~F., Kristiansen, J.~R., Koivisto, T., \& Groeneboom, N.~E. 2007, Mon.
  Not. Roy. Astron. Soc., 382, 793

\bibitem[{Mota {et~al.}(2008{\natexlab{a}})Mota, Pettorino, Robbers, \&
  Wetterich}]{vale}
Mota, D.~F., Pettorino, V., Robbers, G., \& Wetterich, C. 2008{\natexlab{a}},
  Phys. Lett., B663, 160

\bibitem[{Mota \& Shaw(2006)}]{MotaShaw}
Mota, D.~F., \& Shaw, D.~J. 2006, Phys. Rev. Lett., 97, 151102

\bibitem[{{Mota} \& {Shaw}(2007)}]{Mota:2006fz}
{Mota}, D.~F., \& {Shaw}, D.~J. 2007, \prd, 75, 063501

\bibitem[{Mota {et~al.}(2008{\natexlab{b}})Mota, Shaw, \& Silk}]{void}
Mota, D.~F., Shaw, D.~J., \& Silk, J. 2008{\natexlab{b}}, Astrophys. J., 675,
  29

\bibitem[{{Nagata} {et~al.}(2004){Nagata}, {Chiba}, \& {Sugiyama}}]{wmapbound}
{Nagata}, R., {Chiba}, T., \& {Sugiyama}, N. 2004, \prd, 69, 083512

\bibitem[{{Nunes} \& {Mota}(2006)}]{nunes}
{Nunes}, N.~J., \& {Mota}, D.~F. 2006, \mnras, 368, 751

\bibitem[{{Oyaizu} {et~al.}(2008){Oyaizu}, {Lima}, \& {Hu}}]{de8}
{Oyaizu}, H., {Lima}, M., \& {Hu}, W. 2008, \prd, 78, 123524

\bibitem[{{Peebles}(1984)}]{a1}
{Peebles}, P.~J.~E. 1984, \apj, 284, 439

\bibitem[{{Pogosian} \& {Silvestri}(2008)}]{de6}
{Pogosian}, L., \& {Silvestri}, A. 2008, \prd, 77, 023503

\bibitem[{{Sahni} {et~al.}(2003){Sahni}, {Saini}, {Starobinsky}, \&
  {Alam}}]{Sahni:2002fz}
{Sahni}, V., {Saini}, T.~D., {Starobinsky}, A.~A., \& {Alam}, U. 2003, Soviet
  Journal of Experimental and Theoretical Physics Letters, 77, 201

\bibitem[{{Schelpe}(2010)}]{cp2}
{Schelpe}, C.~A.~O. 2010, \prd, 82, 044033

\bibitem[{{Song} {et~al.}(2007{\natexlab{a}}){Song}, {Hu}, \& {Sawicki}}]{de3}
{Song}, Y., {Hu}, W., \& {Sawicki}, I. 2007{\natexlab{a}}, \prd, 75, 044004

\bibitem[{{Song} {et~al.}(2007{\natexlab{b}}){Song}, {Peiris}, \& {Hu}}]{de5}
{Song}, Y., {Peiris}, H., \& {Hu}, W. 2007{\natexlab{b}}, \prd, 76, 063517

\bibitem[{{Steffen} \& {Gammev Collaboration}(2008)}]{local2}
{Steffen}, J.~H., \& {Gammev Collaboration}. 2008, in Identification of Dark
  Matter 2008

\bibitem[{{Tamaki} \& {Tsujikawa}(2008)}]{Tamaki:2008mf}
{Tamaki}, T., \& {Tsujikawa}, S. 2008, \prd, 78, 084028

\bibitem[{{Tatekawa} \& {Tsujikawa}(2008)}]{de7}
{Tatekawa}, T., \& {Tsujikawa}, S. 2008, JCAP, 9, 9

\bibitem[{{Thongkool} {et~al.}(2009){Thongkool}, {Sami}, {Gannouji}, \&
  {Jhingan}}]{Thongkool:2009js}
{Thongkool}, I., {Sami}, M., {Gannouji}, R., \& {Jhingan}, S. 2009, \prd, 80,
  043523

\bibitem[{{Tsujikawa} {et~al.}(2009){Tsujikawa}, {Gannouji}, {Moraes}, \&
  {Polarski}}]{Tsujikawa:2009ku}
{Tsujikawa}, S., {Gannouji}, R., {Moraes}, B., \& {Polarski}, D. 2009, \prd,
  80, 084044

\bibitem[{{Wang} {et~al.}(2000){Wang}, {Caldwell}, {Ostriker}, \&
  {Steinhardt}}]{wang1}
{Wang}, L., {Caldwell}, R.~R., {Ostriker}, J.~P., \& {Steinhardt}, P.~J. 2000,
  \apj, 530, 17

\bibitem[{{Wang} \& {Steinhardt}(1998)}]{wang}
{Wang}, L., \& {Steinhardt}, P.~J. 1998, \apj, 508, 483

\bibitem[{{Wei} \& {Cai}(2005)}]{kchameleon}
{Wei}, H., \& {Cai}, R. 2005, \prd, 71, 043504

\bibitem[{{Will}(1993)}]{Will:93}
{Will}, C.~M. 1993, Theory and experiment in gravitational physics, revised
  edn. (Cambridge University Press)

\bibitem[{{Zhao} {et~al.}(2011){Zhao}, {Li}, \& {Koyama}}]{Li5}
{Zhao}, G., {Li}, B., \& {Koyama}, K. 2011, \prd, 83, 044007

\bibitem[{{Zlatev} {et~al.}(1999){Zlatev}, {Wang}, \& {Steinhardt}}]{zlatev}
{Zlatev}, I., {Wang}, L., \& {Steinhardt}, P.~J. 1999, Physical Review Letters,
  82, 896

\end{thebibliography}
\end{document}